\documentclass[prl,aps,twocolumn,showpacs,floatfix,superscriptaddress]{revtex4-1}
\usepackage{graphicx}
\usepackage{subfigure}
\usepackage{bm}
\usepackage{amsmath}
\usepackage{amssymb}
\usepackage{tabularx}
\usepackage{wasysym}
\usepackage{color}
\usepackage{array}
\usepackage{multirow}
\usepackage{makecell}
\usepackage[version=3]{mhchem} 
\usepackage{float}
\usepackage{color}
\usepackage{hyperref}
\usepackage{lipsum}
\usepackage{ulem}

\makeatletter
\newcommand{\colorcaption}[2][]{%
  \begingroup%
  \renewcommand{\@caption@fignum@sep}{ (color online). }%
  \caption[#1]{#2}%
  \endgroup%
  }
\makeatother

\begin{document}
\title{Electron-hole theory of the effect of quantum nuclei on the X-ray absorption spectra of liquid water}

\author{Zhaoru Sun}
\affiliation{Department of Physics, Temple University, Philadelphia, PA 19122, USA}
\author{Lixin Zheng}
\affiliation{Department of Physics, Temple University, Philadelphia, PA 19122, USA}
\author{Mohan Chen}
\affiliation{Department of Physics, Temple University, Philadelphia, PA 19122, USA}
\author{Michael L. Klein}
\affiliation{Department of Physics, Temple University, Philadelphia, PA 19122, USA}
\affiliation{Department of Chemistry, Temple University, Philadelphia, PA 19122, USA}
\affiliation{Institute for Computational Molecular Science, Temple University, Philadelphia, PA 19122, USA}
\author{Francesco Paesani}
\thanks{$^\ast$Corresponding author. Email:fpaesani@ucsd.edu}
\affiliation{Department of Chemistry and Biochemistry, 
Materials Science and Engineering, San Diego Supercomputer Center, 
University of California, San Diego, La Jolla, California 92093, USA}
\author{Xifan Wu}
\thanks{$^\ast$Corresponding author. Email:xifanwu@temple.edu}
\affiliation{Department of Physics, Temple University, Philadelphia, PA 19122, USA}
\affiliation{Institute for Computational Molecular Science, Temple University, Philadelphia, PA 19122, USA}


\begin{abstract}
Electron-hole excitation theory is used to unveil the role of nuclear quantum effects on the X-ray absorption spectral signatures of water, whose structure is computed via path-integral molecular dynamics with the MB-pol intermolecular potential model. Compared to spectra generated from the classically modeled water, quantum nuclei introduce important effects on the spectra in terms of both the energies and line shapes. Fluctuations due to delocalized protons influence the short-range ordering of the hydrogen bond network via changes in the intramolecular covalence, which broaden the pre-edge spectra. For intermediate-range and long-range ordering, quantum nuclei approach the neighboring oxygen atoms more closely than classical protons, promoting an ``ice-like'' spectral feature with the intensities shifted from the main- to post-edge. Computed spectra are in nearly quantitative agreement with the available experimental data. 
\end{abstract}

\maketitle

The nature of the hydrogen bond (H-bond) network in water continues to occupy a special place at the center of scientific interests \cite{pettersson_lgm_2016,jungwirth_p_2006, bellissent_mc_2016,ball_p_2008,werber_jr_2016,barnett_tp_2005,luzar_a_1996, fecko2003ultrafast, steinel_t_2004, eaves_jd_2005, laage_d_2006,reeves2017electronic}.
Recently, high-resolution X-ray absorption spectroscopy (XAS) has emerged as a powerful experimental technique to probe the structure of water at molecular scale \cite{fransson_t_2016, amann-winkel_k_2016,Wernet995,tse2008PRL,pylkkanen2011temperature,zhang2017multi,nagasaka2017reliable,smith2017soft}.
As allowed by the Franck-Condon principle \cite{bernath2015spectra}, 
electrons leaving a water molecule oxygen-atom core can be described as excited electronic states from an instantaneously frozen snapshot of water at equilibrium. Therefore, the XAS spectrum inherits a unique local signature of its molecular environment, complimentary to traditional X-ray and neutron scattering experiments \cite{soper2008quantum,skinner2013benchmark}.

\textit{Ab initio} theory is essential for an unambiguous interpretation of the underlying electronic structure of water. However, theoretical prediction of the XAS spectrum has posed a major challenge by itself. The computed spectrum is sensitive to the accuracy of methods in both the adopted electronic and molecular theories. Rigorously, the electron-hole excitation theories, such as the Bethe-Saltpeter equation (BSE) \cite{shirley1998ab,rehr2005final,vinson2011bethe,fransson_t_2016}, 
should be applied beyond the unoccupied electronic structure from density functional theory \cite{giulia2006PRLxas,chen2010PRLxas}.
However, the BSE is computational formidable and has not yet widely applied in water. 
This computational burden has been greatly alleviated by the recently introduced approximate BSE solution utilizing localized basis \cite{chen2010PRLxas,kong2012roles,sun2017x}. 
On the other hand, the modeling of water at the molecular level is no less difficult because of the nature of the H-bond, which is a balance of many physical interactions. The directional H- bond, which is the building block of the near-tetrahedral structure in water, is much weaker than covalent or ionic chemical bonds. Furthermore, the H-bond network is influenced by the van der Waals interaction, whose energy is even weaker than H-bond by one order of magnitude \cite{distasio2014individual,wang2011density,chen2017ab,chen2018hydroxide}.
Worse still, nuclear quantum effects (NQEs) should be properly considered due to the low mass of the proton \cite{morrone2008nuclear,ceriotti2016nuclear,galib2017mass}.
Under the influence of NQEs, the more delocalized protons introduce unexpected effects in the H-bond network in addition to the softening of liquid structure \cite{li2011quantum,ceriotti2013nuclear,ceriotti2016nuclear}.
Moreover, NQEs have been shown to have important effects on XAS spectrum \cite{schreck2016isotope,kong2012roles,harada2013selective}.

Notwithstanding the many recent significant advances, challenges remain. 
The precise picture of NQEs on the XAS spectrum at the molecular level remains elusive. Obvious discrepancies are present between theory and experiment even when NQEs are considered in modeling water \cite{kong2012roles}. 
Specifically, the predicted XAS spectrum presented an overestimated spectral intensity in the main-edge and under estimated spectral energies in the post-edge compared to experiments \cite{fransson_t_2016}.
These disagreements reflect the delicate nature of NQEs. 
Delocalized protons (via NQEs) can either strengthen or weaken the H-bond structure, depending on the anharmonicity of the potential energy surface \cite{ceriotti2016nuclear}.
The latter should accurately account for the directionality of the H-bond, as well as dispersion forces arising from many-body effects \cite{cisneros_ga_2016}.
Moreover, compared to the bound exciton described by the pre-edge, the main- and post-edge contributions to the XAS spectrum are resonant excited states, which are sensitive to the intermediate-range and long-range ordering of H-bond network. Hence, a much larger simulation super cell is required to properly describe these delocalized excited states \cite{sun2017x}.

\begin{figure*}[tp]
  \includegraphics[width=17cm]{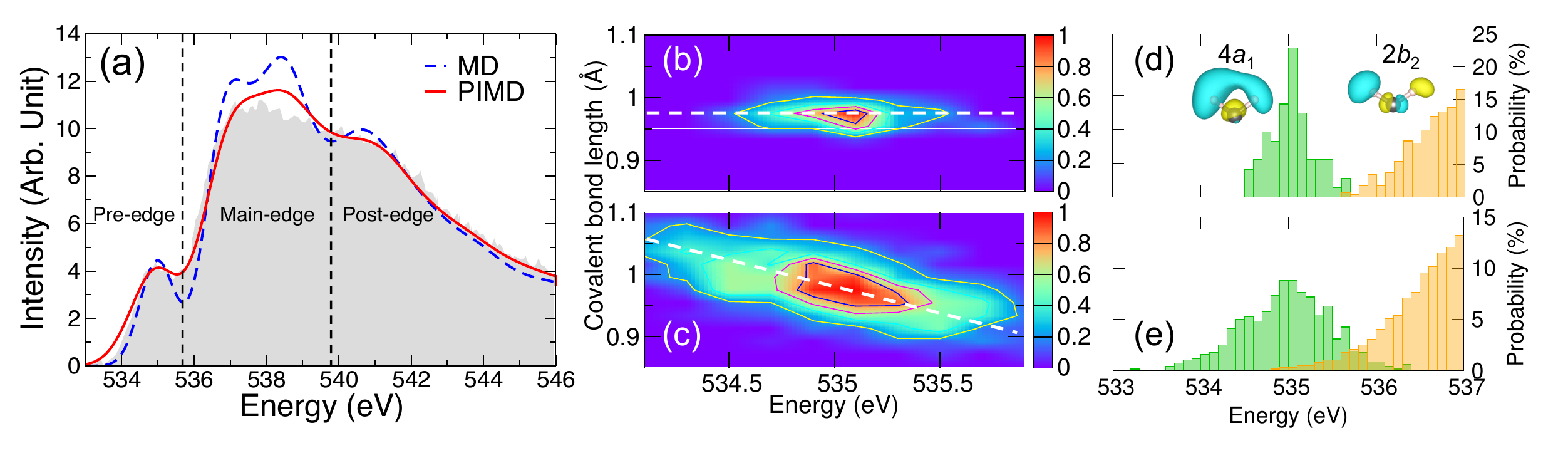}
  \colorcaption{(a) XAS spectra 
from MD (dashed blue) and PIMD (solid red) simulations at 298 K. 
Experimental data \cite{tse2008PRL} is shown in shade.
Joint probability distribution of the covalent bond length 
as a function of the excitation energy for state with $4a_1$ character, from (b) MD and (c) PIMD simulations. 
Probability distribution of the excitation energies for states with $4a_1$ (green) and $2b_2$ (orange) character, 
from (d) MD and (e) PIMD simulations.
}
  \label{figxas}
\end{figure*}

In this letter, we address the above issues. 
NQEs are taken into account via path integral molecular dynamics (PIMD) simulations \cite{paesani_f_2009a}
with the MB-pol many-body potential energy function \cite{babin_v_2013, babin_v_2014, medders_gr_2014}.
Rigorously built upon the many-body expansion of the interaction energy \cite{cisneros_ga_2016},
MB-pol enables the accurate modeling of the properties of water across different 
phases \cite{paesani2016getting, reddy2016accuracy},
from the dimer \cite{babin_v_2013} and small clusters \cite{brown_se_2017}, to liquid water \cite{medders_gr_2014, reddy_sk_2017}, and ice \cite{pham_ch_2017, moberg_dr_2017}.
We use the self-consistent enhanced static Coulomb-hole and screened exchange (COHSEX) approximation \cite{kang2010enhanced} with maximally localized Wannier functions, which greatly enhances the computational efficiency of XAS calculations without compromising the accuracy of the results \cite{09B-Wu,giannozzi2017advanced}.

The theoretical spectrum obtained from PIMD simulation is found to be in excellent agreement with the corresponding experimental data and demonstrates the importance of NQEs for an accurate modeling of XAS spectrum. 
The broadened pre-edge reflects the significantly increased fluctuations in the covalence of the water molecule due to the delocalized protons.
Protons can approach the acceptor oxygen atoms at a much short distance under the influence of NQEs, 
which enhances ``ice-like'' spectral characteristic with a slightly increased intensity of post-edge feature 
with respect to that of the main-edge.
Our XAS spectra at different temperatures suggest slightly larger NQEs on the spectrum at lower temperature, 
which is consistent with our temperature dependent ring-topology analysis of the H-bond network.

All calculations were performed in a super cell containing 128 water molecules
under periodic boundary conditions. 
Molecular configurations of liquid water were extracted from 
classical molecular dynamics (MD) and PIMD simulations carried out with MB-pol
at 270, 298, and 360~K, and were then used 
in calculations of the associated XAS cross sections using 
the enhanced static COHSEX approximation \cite{kang2010enhanced}.
Calculated XAS spectra were adjusted to the same area as in the experimental line shape 
from 533 to 546 eV. 
Also, the pre-edge features were aligned with
the experimental value of 535 eV \cite{chen2010PRLxas,kong2012roles,sun2017x}.
Additional detail about the MD simulations and spectral calculations are given in the 
Supporting Information (SI), 
which includes Refs. \cite{Nose1992,berne1986,reddy2017,ceriotti2012efficient,onida2002electronic}.

\begin{figure*}
  \includegraphics[width=18cm]{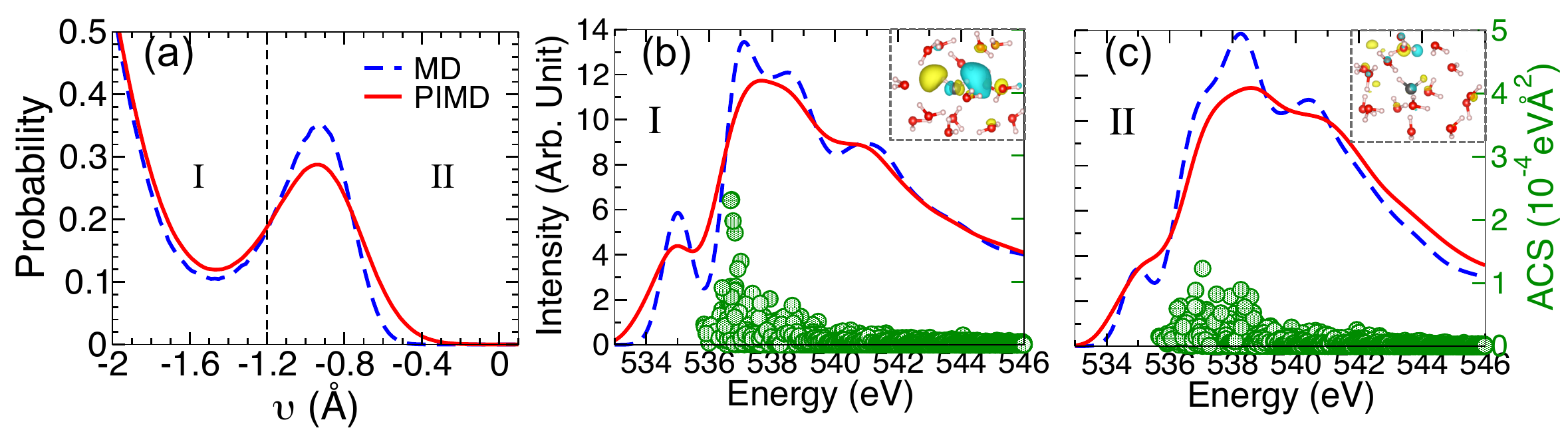}
  \colorcaption{
(a) Distribution of proton-transfer coordinate ($v$) from MD (dashed blue) and 
PIMD (solid red) trajectories. 
XAS spectra are decomposed into contributions from excited water molecules within two $v$ regions (b and c).
Green circles represents absorption cross sections (ACS) of excitation within each $v$ region.
Insets represent excited states with $b_2$ character. The excited oxygen atom is shown in black in the insets.
}
  \label{figmain}
\end{figure*}

The XAS spectra of liquid water calculated
using configurations from MD and PIMD simulations 
at 298~K are shown in Fig. 1(a), along with the corresponding
experimental spectrum \cite{tse2008PRL}.
In the spectrum from MD simulation, the energy of the post-edge is underestimated and both intensities of the main-edge and post-edge are overestimated compared to experiment. In addition, two sub-peaks appear within the main-edge, which are separated from the pre-edge feature by a rather deep minimum that is absent in the experimental spectrum. These drawbacks are corrected in the spectrum from PIMD simulation, which is in nearly quantitative agreement with the corresponding experimental spectrum. 

The pre-edge feature in the XAS spectrum of liquid water
is associated with short-range ordering of the H-bond network \cite{chen2010PRLxas}. 
To provide further molecular-level insights into the relationship between 
the pre-edge feature and the structure of liquid water predicted by both MD and PIMD simulations,
the correlation between the fluctuations of the covalent OH bonds and the pre-edge excitation energies 
is shown in Fig. 1(b and d), with the corresponding distributions of
excitation energies shown in Fig. 1(c and e). 
While both covalent OH bond distributions are centered on 0.97~\AA{} and 535 eV respectively,
the fluctuations are significantly larger in the PIMD simulation due to zero point energy effects.
From this comparison,
it is possible to attribute the differences in the pre-edge feature of the MD and PIMD spectra
to intra-molecular structural changes associated with NQEs, which result in proton delocalization,
and affect the covalent character of the OH bond within each water molecule.

As shown in Fig. 1(d), the pre-edge feature in the spectrum from MD simulation 
is entirely contributed by a bound exciton with $4a_1$ character, 
which is well separated from the main-edge resonant excited states with $2b_2$ character.
In the spectrum from PIMD simulation, the pre-edge excitation energies are statistically linearly correlated 
with the broadening of the covalent OH bond distribution in Fig. 1(c),
which reflects the existence of the distinct quantum effects in liquid water. 
NQEs allow protons to move more easily along the direction of H-bonds, which
thus facilitates the forming (breaking) of H-bonds and the consequent  
decrease (increase) of the covalent character of the associated OH bonds. 
Within this scenario, the excitation energies are reduced (increased) 
due to the increase (decrease) of Coulomb interactions between the protons 
and the electron lone pairs of the oxygens on the neighboring water molecules, 
as demonstrated by the negative correlation obtained in Fig. 1(c). 
NQEs are thus responsible for the broadening of spectrum between 533 and 535.7 eV, 
which leads to nearly quantitative agreement 
with the experimentally observed pre-edge feature.
Due to proton delocalization, the energies of the low-lying resonant excited 
states with $2b_2$ character are also lowered by NQEs. 
As shown in Fig. 1(e), the pre-edge feature in the spectrum from PIMD simulation 
cannot be exclusively attributed to excitations with $4a_1$ character, 
but also contains contributions from excitations with $2b_2$ character,
which results in a smoother separation at $\sim$535.7 eV
between the pre-edge and main-edge features in Fig. 1(a).

While the pre-edge is probing the intra-molecular ordering, 
the main-edge and post-edge features are associated with the resonant excited states 
exploring the H-bond network in the intermediate-range and long-range \cite{sun2017x}.  
Therefore, the NQEs on the XAS spectrum in this energy range are more prone to be affected by 
the inter-molecular structural changes due to the delocalized protons. 
To this end, we resort to the proton-transfer coordinate ($v$) analysis, 
which is sensitive to the inter-molecular proton displacement \cite{ceriotti2013nuclear}. 
The proton-transfer coordinate is defined as, $v=d$(O$-$H)$-d$(O$'-$H), and involves a hydrogen atom H, an oxygen atom O covalently bonded to the H, and a second oxygen atom O$'$ in its first coordination shell, 
with $d$(O-H) representing the distance between oxygen and hydrogen. 
The resulting distributions of $v$ and their comparison between MD and PIMD simulations are presented in Fig. 2(a).

To facilitate the analysis, we further decomposed the distribution of $v$ as well as 
the XAS spectrum into two distinct regions. 
As shown in Fig. 2(a), $v$ in region I is relatively more negative, which indicates that the 
proton is mostly confined within the molecule and well separated from 
the acceptor oxygen atom \cite{ceriotti2013nuclear}. 
The inter-molecular environment in region I therefore depicts a weakly H-bonded structure 
with partial collapse of tetrahedral H-bond network \cite{ceriotti2013nuclear}. 
The above disordered liquid environment encourages strong localizations of excitations in 
main-edge with $b_2$ character \cite{chen2010PRLxas,sun2017x}. 
As a result, the oscillation strengths of the main-edge (labeled by the green circles) in Fig. 2(b) 
are shifted to lower energies compared to the overall XAS spectra in Fig. 1. 
The above can also be seen by the large amplitudes absorption cross sections around 537 eV. 
The increased spectral intensity of the main-edge, in turn, results in the decreased post-edge 
accordingly to satisfy the optical sum rule. 
As a result, the XAS spectrum in region I shows a more ``water-like'' spectral line shape, 
i.e., the intensity of main-edge is significantly more prominent than that of the post-edge. 
It should be noted that this more ``water-like'' spectral shape is the same change in XAS of the water under elevated temperature \cite{chen2010PRLxas}. 
The distributions of $v$ in region I are rather similar between the MD and PIMD simulations 
as shown in Fig. 2(a), except for a slightly increased probability of the later. 
This is also consistent with the more prominent spectral intensities of main-edge than those of post-edge, 
with the ratio of average intensities $\bar{I_M}/\bar{I_P} \approx 1.6$, for both MD and PIMD simulations.

In contrast, the inter-molecular configuration in region II describes a scenario that 
protons are approaching the acceptor oxygen atoms with enhanced probabilities to 
form stronger H-bonds \cite{ceriotti2013nuclear}. 
The strengthened H-bond network in turn encourages the hybridization of the electron excitations 
of $b_2$ character between excited oxygen and surrounding water molecules \cite{chen2010PRLxas,sun2017x}. 
Based on the optical sum rule, the oscillation strengths therefore move from main-edge to post-edge of higher excitation energies as shown in Fig 2(c). 
Keep in mind that the post-edge is more prominent than main-edge in crystalline ice, 
therefore, the XAS spectrum in region II shows a more ``ice-like'' spectral line shape, i.e., 
the intensity of post-edge is increased relatively to that of main-edge \cite{tse2008PRL,chen2010PRLxas}. 
As far as the inter-molecular ordering is concerned in Fig. 2(a), 
relatively large NQEs can be seen here by the promoted configurations with less negative $v$. 
This indicates that the protons have a higher tendency of proton transfer (at $v\sim$ 0) under the influence of NQEs, which is consistent with the findings by Ceriotti \textit{et al} \cite{ceriotti2013nuclear}. 
Not surprisingly, the average intensity ratio between main-edge and post-edge $\bar{I_M}/\bar{I_P}$ yields $\sim$1.3 
in PIMD simulations, which is smaller than the value $\sim$1.5 from MD simulations. 
Hence, the XAS spectrum computed from PIMD simulation displays a more ``ice-like'' spectral line shape 
than those generated from MD simulations. 
In addition, the artificial and over-structured sub-peaks of the main-edge from MD simulation 
are greatly improved in PIMD simulation. 
We attribute the above correction to the proton broadenings at a fixed $v$ as well as the enhanced asymmetric H-bonds 
of excited water molecules under the influence of NQEs \cite{ceriotti2013nuclear,kuhne2013electronic}.

\begin{figure}[tp]
\flushleft
  \includegraphics[width=8.5cm]{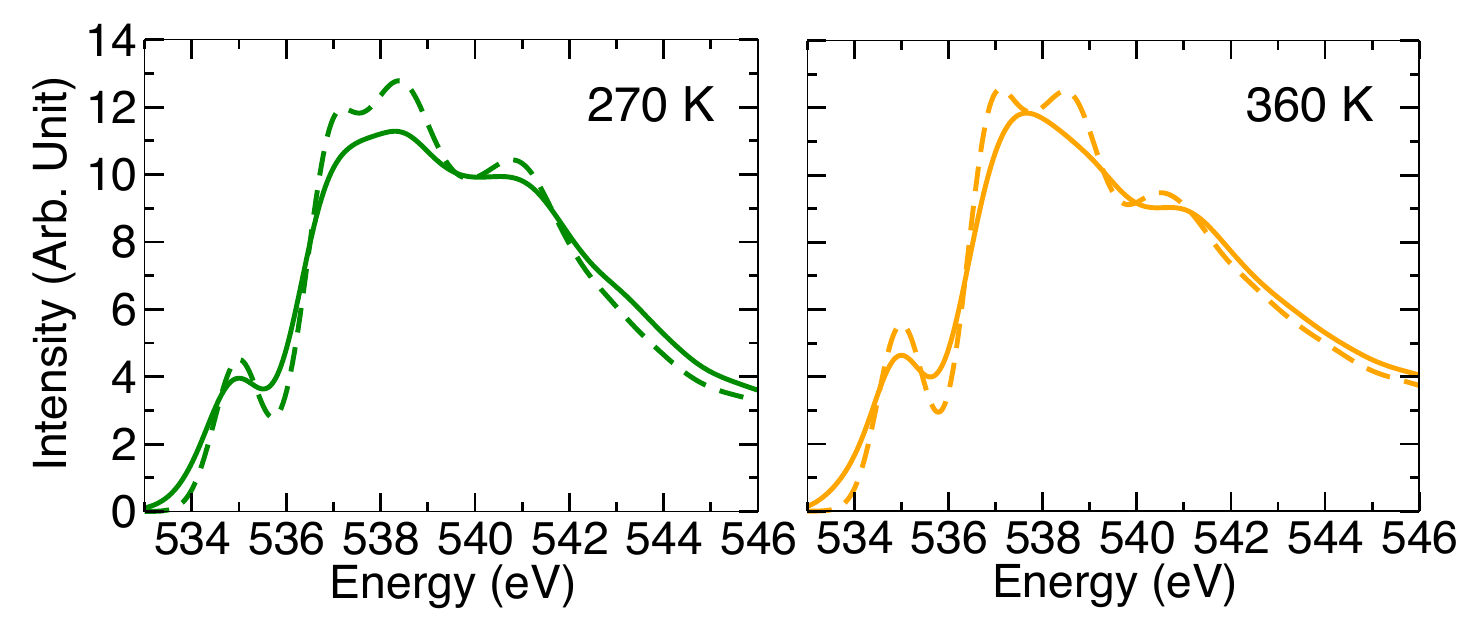}
  \colorcaption{XAS spectra calculated using configurations from MD (dashed) and 
PIMD (solid) simulations at 270 and 360~K.
}
  \label{figtemp}
\end{figure}

To provide further insights into the role NQEs play in determining the local structure
of liquid water, XAS spectra were also calculated at 270 and 360~K as shown in Fig. 3.
Similar to the XAS spectra obtained at 298~K, 
NQEs broaden the pre-edge and smooth the main-edge for all
temperatures. 
Additionally, spectral difference between MD and PIMD simulations at 270~K is slightly
larger than 360~K. 
Due to the different extent of thermal fluctuations and magnitude of associated de Broglie wavelengths,
the local structure of the H-bond network in liquid water changes
significantly as a function of temperature, which is reflected in different spectral line shapes.
Here, we analyze
the topology of the underlying H-bond network in terms of ring structures 
\cite{hassanali2013proton}, whose ring size distributions 
at 270, 298 and 360~K are shown in Fig. 4.
As a reference, crystalline ice I$_h$ is only comprised of six-member rings. 
In liquid water, due to the fluctuating nature of the H-bond network,
there is a distribution of ring sizes, and the number of large-sized
rings increases with increased temperature.
Consistently, the oscillation strength in the XAS spectra gradually 
shifts from the post-edge feature to the main-edge feature with increased temperature, 
which, at the same time, indicates a progressive softening of the liquid structure.
At all temperatures examined in this study, 
NQEs are found to soften the structure of liquid water,
as indicated by slightly broader ring size distributions with 
more large-sized rings obtained from the PIMD simulations. 
As expected based on the dependence of the de Broglie wavelength on temperature, 
the differences between MD and PIMD ring size distributions are more pronounced  
at lower temperature, which is consistent with previous X-ray scattering measurements
\cite{hart2005temperature,hart2006isotope}. 
Such topological H-bond network differences between MD and PIMD at lower and higher temperatures
are also consistent with the spectral change in Fig. 3.

In conclusion, we have reported a systematic analysis of the XAS spectra 
of liquid water calculated at different temperatures from state-of-the-art simulations
of molecular trajectories carried out at both classical and quantum levels using the MB-pol potential energy function.
Our results demonstrate that specific features of the XAS spectra directly report 
the role NQEs play at both the intra- and inter-molecular levels.
These findings reinforce the notion that an accurate representation of the underlying
Born-Oppenheimer potential energy surface and a rigorous account of NQEs are
necessary for a correct description of liquid water.
In future work, it will be interesting to explore whether or not XAS would be able to detect signatures of two-phase coexistence in water \cite{gallo2016water}. 

\begin{figure}[tp]
\flushleft
  \includegraphics[width=8.5cm]{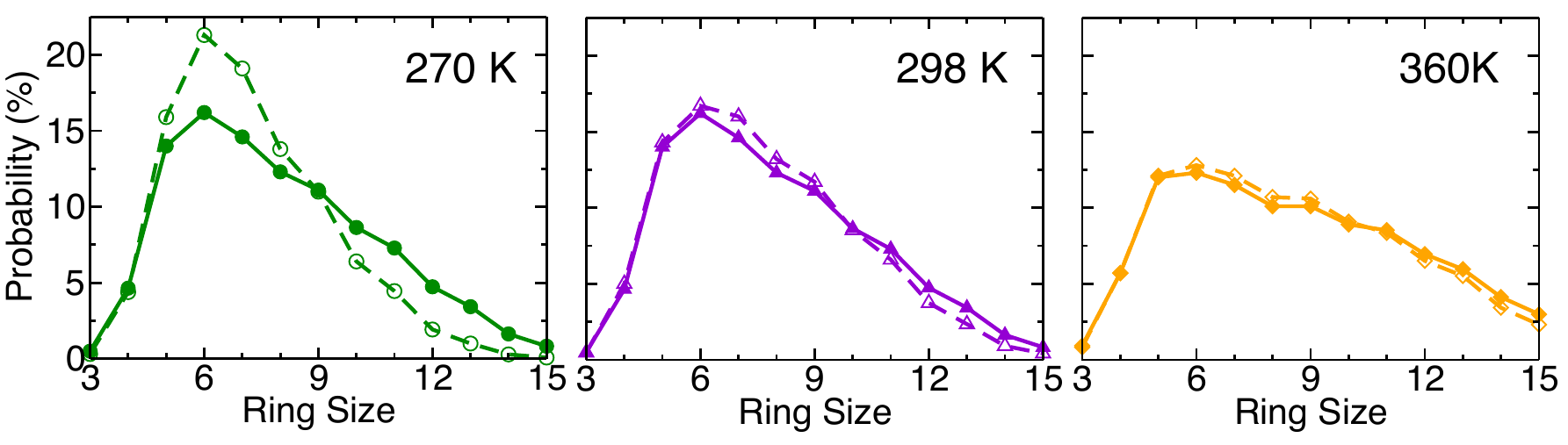}
  \colorcaption{Distributions of ring size at 270, 298, and 360~K obtained 
from MD (dashed) and PIMD (solid) simulations.}
  \label{figring}
\end{figure}

This work was supported by National Science Foundation through
Awards DMR-1552287 (to X.W.) and CHE-1453204 (to F.P.).
This research used resources of the National Energy Research Scientific Computing Center, which is supported by the U.S. Department of Energy (DOE), Office of Science under Contract DEAC02-05CH11231.
The work of Z.S. and M.L.K was supported as part
of the Center for the Computational Design of Functional Layered
Materials, an Energy Frontier Research Center funded by the U.S.
DOE, Office of Science, Basic Energy Sciences under Award DESC0012575.


\begin{thebibliography}{65}%
\makeatletter
\providecommand \@ifxundefined [1]{%
 \@ifx{#1\undefined}
}%
\providecommand \@ifnum [1]{%
 \ifnum #1\expandafter \@firstoftwo
 \else \expandafter \@secondoftwo
 \fi
}%
\providecommand \@ifx [1]{%
 \ifx #1\expandafter \@firstoftwo
 \else \expandafter \@secondoftwo
 \fi
}%
\providecommand \natexlab [1]{#1}%
\providecommand \enquote  [1]{``#1''}%
\providecommand \bibnamefont  [1]{#1}%
\providecommand \bibfnamefont [1]{#1}%
\providecommand \citenamefont [1]{#1}%
\providecommand \href@noop [0]{\@secondoftwo}%
\providecommand \href [0]{\begingroup \@sanitize@url \@href}%
\providecommand \@href[1]{\@@startlink{#1}\@@href}%
\providecommand \@@href[1]{\endgroup#1\@@endlink}%
\providecommand \@sanitize@url [0]{\catcode `\\12\catcode `\$12\catcode
  `\&12\catcode `\#12\catcode `\^12\catcode `\_12\catcode `\%12\relax}%
\providecommand \@@startlink[1]{}%
\providecommand \@@endlink[0]{}%
\providecommand \url  [0]{\begingroup\@sanitize@url \@url }%
\providecommand \@url [1]{\endgroup\@href {#1}{\urlprefix }}%
\providecommand \urlprefix  [0]{URL }%
\providecommand \Eprint [0]{\href }%
\providecommand \doibase [0]{http://dx.doi.org/}%
\providecommand \selectlanguage [0]{\@gobble}%
\providecommand \bibinfo  [0]{\@secondoftwo}%
\providecommand \bibfield  [0]{\@secondoftwo}%
\providecommand \translation [1]{[#1]}%
\providecommand \BibitemOpen [0]{}%
\providecommand \bibitemStop [0]{}%
\providecommand \bibitemNoStop [0]{.\EOS\space}%
\providecommand \EOS [0]{\spacefactor3000\relax}%
\providecommand \BibitemShut  [1]{\csname bibitem#1\endcsname}%
\let\auto@bib@innerbib\@empty
\bibitem [{\citenamefont {Pettersson}\ \emph {et~al.}(2016)\citenamefont
  {Pettersson}, \citenamefont {Henchman},\ and\ \citenamefont
  {Nilsson}}]{pettersson_lgm_2016}%
  \BibitemOpen
  \bibfield  {author} {\bibinfo {author} {\bibfnamefont {L.~G.~M.}\
  \bibnamefont {Pettersson}}, \bibinfo {author} {\bibfnamefont {R.~H.}\
  \bibnamefont {Henchman}}, \ and\ \bibinfo {author} {\bibfnamefont
  {A.}~\bibnamefont {Nilsson}},\ }\href@noop {} {\bibfield  {journal} {\bibinfo
   {journal} {Chem. Rev.}\ }\textbf {\bibinfo {volume} {116}},\ \bibinfo
  {pages} {7459} (\bibinfo {year} {2016})}\BibitemShut {NoStop}%
\bibitem [{\citenamefont {Jungwirth}\ and\ \citenamefont
  {Tobias}(2006)}]{jungwirth_p_2006}%
  \BibitemOpen
  \bibfield  {author} {\bibinfo {author} {\bibfnamefont {P.}~\bibnamefont
  {Jungwirth}}\ and\ \bibinfo {author} {\bibfnamefont {D.~J.}\ \bibnamefont
  {Tobias}},\ }\href@noop {} {\bibfield  {journal} {\bibinfo  {journal} {Chem.
  Rev.}\ }\textbf {\bibinfo {volume} {106}},\ \bibinfo {pages} {1259} (\bibinfo
  {year} {2006})}\BibitemShut {NoStop}%
\bibitem [{\citenamefont {Bellissent-Funel}\ \emph {et~al.}(2016)\citenamefont
  {Bellissent-Funel} \emph {et~al.}}]{bellissent_mc_2016}%
  \BibitemOpen
  \bibfield  {author} {\bibinfo {author} {\bibfnamefont {M.-C.}\ \bibnamefont
  {Bellissent-Funel}} \emph {et~al.},\ }\href@noop {} {\bibfield  {journal}
  {\bibinfo  {journal} {Chem. Rev.}\ }\textbf {\bibinfo {volume} {116}},\
  \bibinfo {pages} {7673} (\bibinfo {year} {2016})}\BibitemShut {NoStop}%
\bibitem [{\citenamefont {Ball}(2008)}]{ball_p_2008}%
  \BibitemOpen
  \bibfield  {author} {\bibinfo {author} {\bibfnamefont {P.}~\bibnamefont
  {Ball}},\ }\href@noop {} {\bibfield  {journal} {\bibinfo  {journal} {Chem.
  Rev.}\ }\textbf {\bibinfo {volume} {108}},\ \bibinfo {pages} {74} (\bibinfo
  {year} {2008})}\BibitemShut {NoStop}%
\bibitem [{\citenamefont {Werber}\ \emph {et~al.}(2016)\citenamefont {Werber},
  \citenamefont {Osuji},\ and\ \citenamefont {Elimelech}}]{werber_jr_2016}%
  \BibitemOpen
  \bibfield  {author} {\bibinfo {author} {\bibfnamefont {J.~R.}\ \bibnamefont
  {Werber}}, \bibinfo {author} {\bibfnamefont {C.~O.}\ \bibnamefont {Osuji}}, \
  and\ \bibinfo {author} {\bibfnamefont {M.}~\bibnamefont {Elimelech}},\
  }\href@noop {} {\bibfield  {journal} {\bibinfo  {journal} {Nat. Rev. Mater.}\
  }\textbf {\bibinfo {volume} {1}},\ \bibinfo {pages} {16018} (\bibinfo {year}
  {2016})}\BibitemShut {NoStop}%
\bibitem [{\citenamefont {Barnett}\ \emph {et~al.}(2005)\citenamefont
  {Barnett}, \citenamefont {Adam},\ and\ \citenamefont
  {Lettenmaier}}]{barnett_tp_2005}%
  \BibitemOpen
  \bibfield  {author} {\bibinfo {author} {\bibfnamefont {T.~P.}\ \bibnamefont
  {Barnett}}, \bibinfo {author} {\bibfnamefont {J.~C.}\ \bibnamefont {Adam}}, \
  and\ \bibinfo {author} {\bibfnamefont {D.~P.}\ \bibnamefont {Lettenmaier}},\
  }\href@noop {} {\bibfield  {journal} {\bibinfo  {journal} {Nature}\ }\textbf
  {\bibinfo {volume} {438}},\ \bibinfo {pages} {303} (\bibinfo {year}
  {2005})}\BibitemShut {NoStop}%
\bibitem [{\citenamefont {Luzar}\ and\ \citenamefont
  {Chandler}(1996)}]{luzar_a_1996}%
  \BibitemOpen
  \bibfield  {author} {\bibinfo {author} {\bibfnamefont {A.}~\bibnamefont
  {Luzar}}\ and\ \bibinfo {author} {\bibfnamefont {D.}~\bibnamefont
  {Chandler}},\ }\href@noop {} {\bibfield  {journal} {\bibinfo  {journal}
  {Nature}\ }\textbf {\bibinfo {volume} {379}},\ \bibinfo {pages} {55}
  (\bibinfo {year} {1996})}\BibitemShut {NoStop}%
\bibitem [{\citenamefont {Fecko}\ \emph {et~al.}(2003)\citenamefont {Fecko},
  \citenamefont {Eaves}, \citenamefont {Loparo}, \citenamefont {Tokmakoff},\
  and\ \citenamefont {Geissler}}]{fecko2003ultrafast}%
  \BibitemOpen
  \bibfield  {author} {\bibinfo {author} {\bibfnamefont {C.}~\bibnamefont
  {Fecko}}, \bibinfo {author} {\bibfnamefont {J.}~\bibnamefont {Eaves}},
  \bibinfo {author} {\bibfnamefont {J.}~\bibnamefont {Loparo}}, \bibinfo
  {author} {\bibfnamefont {A.}~\bibnamefont {Tokmakoff}}, \ and\ \bibinfo
  {author} {\bibfnamefont {P.}~\bibnamefont {Geissler}},\ }\href@noop {}
  {\bibfield  {journal} {\bibinfo  {journal} {Science}\ }\textbf {\bibinfo
  {volume} {301}},\ \bibinfo {pages} {1698} (\bibinfo {year}
  {2003})}\BibitemShut {NoStop}%
\bibitem [{\citenamefont {Steinel}\ \emph {et~al.}(2004)\citenamefont
  {Steinel}, \citenamefont {Asbury}, \citenamefont {Zheng},\ and\ \citenamefont
  {Fayer}}]{steinel_t_2004}%
  \BibitemOpen
  \bibfield  {author} {\bibinfo {author} {\bibfnamefont {T.}~\bibnamefont
  {Steinel}}, \bibinfo {author} {\bibfnamefont {J.~B.}\ \bibnamefont {Asbury}},
  \bibinfo {author} {\bibfnamefont {J.}~\bibnamefont {Zheng}}, \ and\ \bibinfo
  {author} {\bibfnamefont {M.}~\bibnamefont {Fayer}},\ }\href@noop {}
  {\bibfield  {journal} {\bibinfo  {journal} {J. Phys. Chem. A}\ }\textbf
  {\bibinfo {volume} {108}},\ \bibinfo {pages} {10957} (\bibinfo {year}
  {2004})}\BibitemShut {NoStop}%
\bibitem [{\citenamefont {Eaves}\ \emph {et~al.}(2005)\citenamefont {Eaves},
  \citenamefont {Loparo}, \citenamefont {Fecko}, \citenamefont {Roberts},
  \citenamefont {Tokmakoff},\ and\ \citenamefont {Geissler}}]{eaves_jd_2005}%
  \BibitemOpen
  \bibfield  {author} {\bibinfo {author} {\bibfnamefont {J.}~\bibnamefont
  {Eaves}}, \bibinfo {author} {\bibfnamefont {J.}~\bibnamefont {Loparo}},
  \bibinfo {author} {\bibfnamefont {C.~J.}\ \bibnamefont {Fecko}}, \bibinfo
  {author} {\bibfnamefont {S.}~\bibnamefont {Roberts}}, \bibinfo {author}
  {\bibfnamefont {A.}~\bibnamefont {Tokmakoff}}, \ and\ \bibinfo {author}
  {\bibfnamefont {P.}~\bibnamefont {Geissler}},\ }\href@noop {} {\bibfield
  {journal} {\bibinfo  {journal} {Proc. Natl. Acad. Sci. U.S.A.}\ }\textbf
  {\bibinfo {volume} {102}},\ \bibinfo {pages} {13019} (\bibinfo {year}
  {2005})}\BibitemShut {NoStop}%
\bibitem [{\citenamefont {Laage}\ and\ \citenamefont
  {Hynes}(2006)}]{laage_d_2006}%
  \BibitemOpen
  \bibfield  {author} {\bibinfo {author} {\bibfnamefont {D.}~\bibnamefont
  {Laage}}\ and\ \bibinfo {author} {\bibfnamefont {J.~T.}\ \bibnamefont
  {Hynes}},\ }\href@noop {} {\bibfield  {journal} {\bibinfo  {journal}
  {Science}\ }\textbf {\bibinfo {volume} {311}},\ \bibinfo {pages} {832}
  (\bibinfo {year} {2006})}\BibitemShut {NoStop}%
\bibitem [{\citenamefont {Reeves}\ and\ \citenamefont
  {Kanai}(2017)}]{reeves2017electronic}%
  \BibitemOpen
  \bibfield  {author} {\bibinfo {author} {\bibfnamefont {K.~G.}\ \bibnamefont
  {Reeves}}\ and\ \bibinfo {author} {\bibfnamefont {Y.}~\bibnamefont {Kanai}},\
  }\href@noop {} {\bibfield  {journal} {\bibinfo  {journal} {Sci. Rep.}\
  }\textbf {\bibinfo {volume} {7}},\ \bibinfo {pages} {40379} (\bibinfo {year}
  {2017})}\BibitemShut {NoStop}%
\bibitem [{\citenamefont {Fransson}\ \emph {et~al.}(2016)\citenamefont
  {Fransson} \emph {et~al.}}]{fransson_t_2016}%
  \BibitemOpen
  \bibfield  {author} {\bibinfo {author} {\bibfnamefont {T.}~\bibnamefont
  {Fransson}} \emph {et~al.},\ }\href@noop {} {\bibfield  {journal} {\bibinfo
  {journal} {Chem. Rev.}\ }\textbf {\bibinfo {volume} {116}},\ \bibinfo {pages}
  {7551} (\bibinfo {year} {2016})}\BibitemShut {NoStop}%
\bibitem [{\citenamefont {Amann-Winkel}\ \emph {et~al.}(2016)\citenamefont
  {Amann-Winkel}, \citenamefont {Bellissent-Funel}, \citenamefont {Bove},
  \citenamefont {Loerting}, \citenamefont {Nilsson}, \citenamefont {Paciaroni},
  \citenamefont {Schlesinger},\ and\ \citenamefont
  {Skinner}}]{amann-winkel_k_2016}%
  \BibitemOpen
  \bibfield  {author} {\bibinfo {author} {\bibfnamefont {K.}~\bibnamefont
  {Amann-Winkel}}, \bibinfo {author} {\bibfnamefont {M.-C.}\ \bibnamefont
  {Bellissent-Funel}}, \bibinfo {author} {\bibfnamefont {L.~E.}\ \bibnamefont
  {Bove}}, \bibinfo {author} {\bibfnamefont {T.}~\bibnamefont {Loerting}},
  \bibinfo {author} {\bibfnamefont {A.}~\bibnamefont {Nilsson}}, \bibinfo
  {author} {\bibfnamefont {A.}~\bibnamefont {Paciaroni}}, \bibinfo {author}
  {\bibfnamefont {D.}~\bibnamefont {Schlesinger}}, \ and\ \bibinfo {author}
  {\bibfnamefont {L.}~\bibnamefont {Skinner}},\ }\href@noop {} {\bibfield
  {journal} {\bibinfo  {journal} {Chem. Rev.}\ }\textbf {\bibinfo {volume}
  {116}},\ \bibinfo {pages} {7570} (\bibinfo {year} {2016})}\BibitemShut
  {NoStop}%
\bibitem [{\citenamefont {Wernet}\ \emph {et~al.}(2004)\citenamefont {Wernet}
  \emph {et~al.}}]{Wernet995}%
  \BibitemOpen
  \bibfield  {author} {\bibinfo {author} {\bibfnamefont {P.}~\bibnamefont
  {Wernet}} \emph {et~al.},\ }\href@noop {} {\bibfield  {journal} {\bibinfo
  {journal} {Science}\ }\textbf {\bibinfo {volume} {304}},\ \bibinfo {pages}
  {995} (\bibinfo {year} {2004})}\BibitemShut {NoStop}%
\bibitem [{\citenamefont {John}\ \emph {et~al.}(2008)\citenamefont {John},
  \citenamefont {Shaw}, \citenamefont {Klug}, \citenamefont {Patchkovskii},
  \citenamefont {Vank{\'o}}, \citenamefont {Monaco},\ and\ \citenamefont
  {Krisch}}]{tse2008PRL}%
  \BibitemOpen
  \bibfield  {author} {\bibinfo {author} {\bibfnamefont {S.~T.}\ \bibnamefont
  {John}}, \bibinfo {author} {\bibfnamefont {D.~M.}\ \bibnamefont {Shaw}},
  \bibinfo {author} {\bibfnamefont {D.~D.}\ \bibnamefont {Klug}}, \bibinfo
  {author} {\bibfnamefont {S.}~\bibnamefont {Patchkovskii}}, \bibinfo {author}
  {\bibfnamefont {G.}~\bibnamefont {Vank{\'o}}}, \bibinfo {author}
  {\bibfnamefont {G.}~\bibnamefont {Monaco}}, \ and\ \bibinfo {author}
  {\bibfnamefont {M.}~\bibnamefont {Krisch}},\ }\href@noop {} {\bibfield
  {journal} {\bibinfo  {journal} {Phys. Rev. Lett.}\ }\textbf {\bibinfo
  {volume} {100}},\ \bibinfo {pages} {095502} (\bibinfo {year}
  {2008})}\BibitemShut {NoStop}%
\bibitem [{\citenamefont {Pylkkanen}\ \emph {et~al.}(2011)\citenamefont
  {Pylkkanen}, \citenamefont {Sakko}, \citenamefont {Hakala}, \citenamefont
  {Hamalainen}, \citenamefont {Monaco},\ and\ \citenamefont
  {Huotari}}]{pylkkanen2011temperature}%
  \BibitemOpen
  \bibfield  {author} {\bibinfo {author} {\bibfnamefont {T.}~\bibnamefont
  {Pylkkanen}}, \bibinfo {author} {\bibfnamefont {A.}~\bibnamefont {Sakko}},
  \bibinfo {author} {\bibfnamefont {M.}~\bibnamefont {Hakala}}, \bibinfo
  {author} {\bibfnamefont {K.}~\bibnamefont {Hamalainen}}, \bibinfo {author}
  {\bibfnamefont {G.}~\bibnamefont {Monaco}}, \ and\ \bibinfo {author}
  {\bibfnamefont {S.}~\bibnamefont {Huotari}},\ }\href@noop {} {\bibfield
  {journal} {\bibinfo  {journal} {J. Phys. Chem. B}\ }\textbf {\bibinfo
  {volume} {115}},\ \bibinfo {pages} {14544} (\bibinfo {year}
  {2011})}\BibitemShut {NoStop}%
\bibitem [{\citenamefont {Zhang}\ \emph {et~al.}(2017)\citenamefont {Zhang},
  \citenamefont {Topsakal}, \citenamefont {Cama}, \citenamefont {Pelliccione},
  \citenamefont {Zhao}, \citenamefont {Ehrlich}, \citenamefont {Wu},
  \citenamefont {Zhu}, \citenamefont {Frenkel}, \citenamefont {Takeuchi},
  \citenamefont {Takeuchi}, \citenamefont {Marschilok}, \citenamefont {Lu},\
  and\ \citenamefont {Wang}}]{zhang2017multi}%
  \BibitemOpen
  \bibfield  {author} {\bibinfo {author} {\bibfnamefont {W.}~\bibnamefont
  {Zhang}}, \bibinfo {author} {\bibfnamefont {M.}~\bibnamefont {Topsakal}},
  \bibinfo {author} {\bibfnamefont {C.}~\bibnamefont {Cama}}, \bibinfo {author}
  {\bibfnamefont {C.~J.}\ \bibnamefont {Pelliccione}}, \bibinfo {author}
  {\bibfnamefont {H.}~\bibnamefont {Zhao}}, \bibinfo {author} {\bibfnamefont
  {S.}~\bibnamefont {Ehrlich}}, \bibinfo {author} {\bibfnamefont
  {L.}~\bibnamefont {Wu}}, \bibinfo {author} {\bibfnamefont {Y.}~\bibnamefont
  {Zhu}}, \bibinfo {author} {\bibfnamefont {A.~I.}\ \bibnamefont {Frenkel}},
  \bibinfo {author} {\bibfnamefont {K.~J.}\ \bibnamefont {Takeuchi}}, \bibinfo
  {author} {\bibfnamefont {E.~S.}\ \bibnamefont {Takeuchi}}, \bibinfo {author}
  {\bibfnamefont {A.~C.}\ \bibnamefont {Marschilok}}, \bibinfo {author}
  {\bibfnamefont {D.}~\bibnamefont {Lu}}, \ and\ \bibinfo {author}
  {\bibfnamefont {F.}~\bibnamefont {Wang}},\ }\href@noop {} {\bibfield
  {journal} {\bibinfo  {journal} {J. Am. Chem. Soc.}\ }\textbf {\bibinfo
  {volume} {139}},\ \bibinfo {pages} {16591} (\bibinfo {year}
  {2017})}\BibitemShut {NoStop}%
\bibitem [{\citenamefont {Nagasaka}\ \emph {et~al.}(2017)\citenamefont
  {Nagasaka}, \citenamefont {Yuzawa}, \citenamefont {Horigome},\ and\
  \citenamefont {Kosugi}}]{nagasaka2017reliable}%
  \BibitemOpen
  \bibfield  {author} {\bibinfo {author} {\bibfnamefont {M.}~\bibnamefont
  {Nagasaka}}, \bibinfo {author} {\bibfnamefont {H.}~\bibnamefont {Yuzawa}},
  \bibinfo {author} {\bibfnamefont {T.}~\bibnamefont {Horigome}}, \ and\
  \bibinfo {author} {\bibfnamefont {N.}~\bibnamefont {Kosugi}},\ }\href@noop {}
  {\bibfield  {journal} {\bibinfo  {journal} {J. Electron. Spectrosc. Relat.
  Phenom.}\ } (\bibinfo {year} {2017})}\BibitemShut {NoStop}%
\bibitem [{\citenamefont {Smith}\ and\ \citenamefont
  {Saykally}(2017)}]{smith2017soft}%
  \BibitemOpen
  \bibfield  {author} {\bibinfo {author} {\bibfnamefont {J.~W.}\ \bibnamefont
  {Smith}}\ and\ \bibinfo {author} {\bibfnamefont {R.~J.}\ \bibnamefont
  {Saykally}},\ }\href@noop {} {\bibfield  {journal} {\bibinfo  {journal}
  {Chem. Rev.}\ }\textbf {\bibinfo {volume} {117}},\ \bibinfo {pages} {13909}
  (\bibinfo {year} {2017})}\BibitemShut {NoStop}%
\bibitem [{\citenamefont {Bernath}(2015)}]{bernath2015spectra}%
  \BibitemOpen
  \bibfield  {author} {\bibinfo {author} {\bibfnamefont {P.~F.}\ \bibnamefont
  {Bernath}},\ }\href@noop {} {\emph {\bibinfo {title} {Spectra of atoms and
  molecules}}}\ (\bibinfo  {publisher} {Oxford university press},\ \bibinfo
  {year} {2015})\BibitemShut {NoStop}%
\bibitem [{\citenamefont {Soper}\ and\ \citenamefont
  {Benmore}(2008)}]{soper2008quantum}%
  \BibitemOpen
  \bibfield  {author} {\bibinfo {author} {\bibfnamefont {A.~K.}\ \bibnamefont
  {Soper}}\ and\ \bibinfo {author} {\bibfnamefont {C.~J.}\ \bibnamefont
  {Benmore}},\ }\href@noop {} {\bibfield  {journal} {\bibinfo  {journal} {Phys.
  Rev. Lett.}\ }\textbf {\bibinfo {volume} {101}},\ \bibinfo {pages} {065502}
  (\bibinfo {year} {2008})}\BibitemShut {NoStop}%
\bibitem [{\citenamefont {Skinner}\ \emph {et~al.}(2013)\citenamefont
  {Skinner}, \citenamefont {Huang}, \citenamefont {Schelesinger}, \citenamefont
  {Pettersson}, \citenamefont {Nilsson},\ and\ \citenamefont
  {Benmore}}]{skinner2013benchmark}%
  \BibitemOpen
  \bibfield  {author} {\bibinfo {author} {\bibfnamefont {L.~B.}\ \bibnamefont
  {Skinner}}, \bibinfo {author} {\bibfnamefont {C.}~\bibnamefont {Huang}},
  \bibinfo {author} {\bibfnamefont {D.}~\bibnamefont {Schelesinger}}, \bibinfo
  {author} {\bibfnamefont {L.~G.}\ \bibnamefont {Pettersson}}, \bibinfo
  {author} {\bibfnamefont {A.}~\bibnamefont {Nilsson}}, \ and\ \bibinfo
  {author} {\bibfnamefont {C.~J.}\ \bibnamefont {Benmore}},\ }\href@noop {}
  {\bibfield  {journal} {\bibinfo  {journal} {J. Chem. Phys.}\ }\textbf
  {\bibinfo {volume} {138}},\ \bibinfo {pages} {074506} (\bibinfo {year}
  {2013})}\BibitemShut {NoStop}%
\bibitem [{\citenamefont {Shirley}(1998)}]{shirley1998ab}%
  \BibitemOpen
  \bibfield  {author} {\bibinfo {author} {\bibfnamefont {E.~L.}\ \bibnamefont
  {Shirley}},\ }\href@noop {} {\bibfield  {journal} {\bibinfo  {journal} {Phys.
  Rev. Lett.}\ }\textbf {\bibinfo {volume} {80}},\ \bibinfo {pages} {794}
  (\bibinfo {year} {1998})}\BibitemShut {NoStop}%
\bibitem [{\citenamefont {Rehr}\ \emph {et~al.}(2005)\citenamefont {Rehr},
  \citenamefont {Soininen},\ and\ \citenamefont {Shirley}}]{rehr2005final}%
  \BibitemOpen
  \bibfield  {author} {\bibinfo {author} {\bibfnamefont {J.}~\bibnamefont
  {Rehr}}, \bibinfo {author} {\bibfnamefont {J.}~\bibnamefont {Soininen}}, \
  and\ \bibinfo {author} {\bibfnamefont {E.~L.}\ \bibnamefont {Shirley}},\
  }\href@noop {} {\bibfield  {journal} {\bibinfo  {journal} {Phys. Scr.}\
  }\textbf {\bibinfo {volume} {2005}},\ \bibinfo {pages} {207} (\bibinfo {year}
  {2005})}\BibitemShut {NoStop}%
\bibitem [{\citenamefont {Vinson}\ \emph {et~al.}(2011)\citenamefont {Vinson},
  \citenamefont {Rehr}, \citenamefont {Kas},\ and\ \citenamefont
  {Shirley}}]{vinson2011bethe}%
  \BibitemOpen
  \bibfield  {author} {\bibinfo {author} {\bibfnamefont {J.}~\bibnamefont
  {Vinson}}, \bibinfo {author} {\bibfnamefont {J.~J.}\ \bibnamefont {Rehr}},
  \bibinfo {author} {\bibfnamefont {J.~J.}\ \bibnamefont {Kas}}, \ and\
  \bibinfo {author} {\bibfnamefont {E.~L.}\ \bibnamefont {Shirley}},\
  }\href@noop {} {\bibfield  {journal} {\bibinfo  {journal} {Phys. Rev. B}\
  }\textbf {\bibinfo {volume} {83}},\ \bibinfo {pages} {115106} (\bibinfo
  {year} {2011})}\BibitemShut {NoStop}%
\bibitem [{\citenamefont {Prendergast}\ and\ \citenamefont
  {Galli}(2006)}]{giulia2006PRLxas}%
  \BibitemOpen
  \bibfield  {author} {\bibinfo {author} {\bibfnamefont {D.}~\bibnamefont
  {Prendergast}}\ and\ \bibinfo {author} {\bibfnamefont {G.}~\bibnamefont
  {Galli}},\ }\href {\doibase 10.1103/PhysRevLett.96.215502} {\bibfield
  {journal} {\bibinfo  {journal} {Phys. Rev. Lett.}\ }\textbf {\bibinfo
  {volume} {96}},\ \bibinfo {pages} {215502} (\bibinfo {year}
  {2006})}\BibitemShut {NoStop}%
\bibitem [{\citenamefont {Chen}\ \emph {et~al.}(2010)\citenamefont {Chen},
  \citenamefont {Wu},\ and\ \citenamefont {Car}}]{chen2010PRLxas}%
  \BibitemOpen
  \bibfield  {author} {\bibinfo {author} {\bibfnamefont {W.}~\bibnamefont
  {Chen}}, \bibinfo {author} {\bibfnamefont {X.}~\bibnamefont {Wu}}, \ and\
  \bibinfo {author} {\bibfnamefont {R.}~\bibnamefont {Car}},\ }\href@noop {}
  {\bibfield  {journal} {\bibinfo  {journal} {Phys. Rev. Lett.}\ }\textbf
  {\bibinfo {volume} {105}},\ \bibinfo {pages} {017802} (\bibinfo {year}
  {2010})}\BibitemShut {NoStop}%
\bibitem [{\citenamefont {Kong}\ \emph {et~al.}(2012)\citenamefont {Kong},
  \citenamefont {Wu},\ and\ \citenamefont {Car}}]{kong2012roles}%
  \BibitemOpen
  \bibfield  {author} {\bibinfo {author} {\bibfnamefont {L.}~\bibnamefont
  {Kong}}, \bibinfo {author} {\bibfnamefont {X.}~\bibnamefont {Wu}}, \ and\
  \bibinfo {author} {\bibfnamefont {R.}~\bibnamefont {Car}},\ }\href@noop {}
  {\bibfield  {journal} {\bibinfo  {journal} {Phys. Rev. B}\ }\textbf {\bibinfo
  {volume} {86}},\ \bibinfo {pages} {134203} (\bibinfo {year}
  {2012})}\BibitemShut {NoStop}%
\bibitem [{\citenamefont {Sun}\ \emph {et~al.}(2017)\citenamefont {Sun} \emph
  {et~al.}}]{sun2017x}%
  \BibitemOpen
  \bibfield  {author} {\bibinfo {author} {\bibfnamefont {Z.}~\bibnamefont
  {Sun}} \emph {et~al.},\ }\href@noop {} {\bibfield  {journal} {\bibinfo
  {journal} {Phys. Rev. B}\ }\textbf {\bibinfo {volume} {96}},\ \bibinfo
  {pages} {104202} (\bibinfo {year} {2017})}\BibitemShut {NoStop}%
\bibitem [{\citenamefont {DiStasio~Jr}\ \emph {et~al.}(2014)\citenamefont
  {DiStasio~Jr}, \citenamefont {Santra}, \citenamefont {Li}, \citenamefont
  {Wu},\ and\ \citenamefont {Car}}]{distasio2014individual}%
  \BibitemOpen
  \bibfield  {author} {\bibinfo {author} {\bibfnamefont {R.~A.}\ \bibnamefont
  {DiStasio~Jr}}, \bibinfo {author} {\bibfnamefont {B.}~\bibnamefont {Santra}},
  \bibinfo {author} {\bibfnamefont {Z.}~\bibnamefont {Li}}, \bibinfo {author}
  {\bibfnamefont {X.}~\bibnamefont {Wu}}, \ and\ \bibinfo {author}
  {\bibfnamefont {R.}~\bibnamefont {Car}},\ }\href@noop {} {\bibfield
  {journal} {\bibinfo  {journal} {J. Chem. Phys.}\ }\textbf {\bibinfo {volume}
  {141}},\ \bibinfo {pages} {084502} (\bibinfo {year} {2014})}\BibitemShut
  {NoStop}%
\bibitem [{\citenamefont {Wang}\ \emph {et~al.}(2011)\citenamefont {Wang},
  \citenamefont {Rom{\'a}n-P{\'e}rez}, \citenamefont {Soler}, \citenamefont
  {Artacho},\ and\ \citenamefont {Fern{\'a}ndez-Serra}}]{wang2011density}%
  \BibitemOpen
  \bibfield  {author} {\bibinfo {author} {\bibfnamefont {J.}~\bibnamefont
  {Wang}}, \bibinfo {author} {\bibfnamefont {G.}~\bibnamefont
  {Rom{\'a}n-P{\'e}rez}}, \bibinfo {author} {\bibfnamefont {J.~M.}\
  \bibnamefont {Soler}}, \bibinfo {author} {\bibfnamefont {E.}~\bibnamefont
  {Artacho}}, \ and\ \bibinfo {author} {\bibfnamefont {M.-V.}\ \bibnamefont
  {Fern{\'a}ndez-Serra}},\ }\href@noop {} {\bibfield  {journal} {\bibinfo
  {journal} {J. Chem. Phys.}\ }\textbf {\bibinfo {volume} {134}},\ \bibinfo
  {pages} {024516} (\bibinfo {year} {2011})}\BibitemShut {NoStop}%
\bibitem [{\citenamefont {Chen}\ \emph {et~al.}(2017)\citenamefont {Chen} \emph
  {et~al.}}]{chen2017ab}%
  \BibitemOpen
  \bibfield  {author} {\bibinfo {author} {\bibfnamefont {M.}~\bibnamefont
  {Chen}} \emph {et~al.},\ }\href@noop {} {\bibfield  {journal} {\bibinfo
  {journal} {Proc. Natl. Acad. Sci. U.S.A.}\ }\textbf {\bibinfo {volume}
  {114}},\ \bibinfo {pages} {10846} (\bibinfo {year} {2017})}\BibitemShut
  {NoStop}%
\bibitem [{\citenamefont {Chen}\ \emph {et~al.}(2018)\citenamefont {Chen} \emph
  {et~al.}}]{chen2018hydroxide}%
  \BibitemOpen
  \bibfield  {author} {\bibinfo {author} {\bibfnamefont {M.}~\bibnamefont
  {Chen}} \emph {et~al.},\ }\href@noop {} {\bibfield  {journal} {\bibinfo
  {journal} {Nat. Chem.}\ }\textbf {\bibinfo {volume} {10}},\ \bibinfo {pages}
  {413} (\bibinfo {year} {2018})}\BibitemShut {NoStop}%
\bibitem [{\citenamefont {Morrone}\ and\ \citenamefont
  {Car}(2008)}]{morrone2008nuclear}%
  \BibitemOpen
  \bibfield  {author} {\bibinfo {author} {\bibfnamefont {J.~A.}\ \bibnamefont
  {Morrone}}\ and\ \bibinfo {author} {\bibfnamefont {R.}~\bibnamefont {Car}},\
  }\href@noop {} {\bibfield  {journal} {\bibinfo  {journal} {Phys. Rev. Lett.}\
  }\textbf {\bibinfo {volume} {101}},\ \bibinfo {pages} {017801} (\bibinfo
  {year} {2008})}\BibitemShut {NoStop}%
\bibitem [{\citenamefont {Ceriotti}\ \emph {et~al.}(2016)\citenamefont
  {Ceriotti} \emph {et~al.}}]{ceriotti2016nuclear}%
  \BibitemOpen
  \bibfield  {author} {\bibinfo {author} {\bibfnamefont {M.}~\bibnamefont
  {Ceriotti}} \emph {et~al.},\ }\href@noop {} {\bibfield  {journal} {\bibinfo
  {journal} {Chem. Rev.}\ }\textbf {\bibinfo {volume} {116}},\ \bibinfo {pages}
  {7529} (\bibinfo {year} {2016})}\BibitemShut {NoStop}%
\bibitem [{\citenamefont {Galib}\ \emph {et~al.}(2017)\citenamefont {Galib},
  \citenamefont {Duignan}, \citenamefont {Misteli}, \citenamefont {Baer},
  \citenamefont {Schenter}, \citenamefont {Hutter},\ and\ \citenamefont
  {Mundy}}]{galib2017mass}%
  \BibitemOpen
  \bibfield  {author} {\bibinfo {author} {\bibfnamefont {M.}~\bibnamefont
  {Galib}}, \bibinfo {author} {\bibfnamefont {T.~T.}\ \bibnamefont {Duignan}},
  \bibinfo {author} {\bibfnamefont {Y.}~\bibnamefont {Misteli}}, \bibinfo
  {author} {\bibfnamefont {M.~D.}\ \bibnamefont {Baer}}, \bibinfo {author}
  {\bibfnamefont {G.~K.}\ \bibnamefont {Schenter}}, \bibinfo {author}
  {\bibfnamefont {J.}~\bibnamefont {Hutter}}, \ and\ \bibinfo {author}
  {\bibfnamefont {C.~J.}\ \bibnamefont {Mundy}},\ }\href@noop {} {\bibfield
  {journal} {\bibinfo  {journal} {J. Chem. Phys.}\ }\textbf {\bibinfo {volume}
  {146}},\ \bibinfo {pages} {244501} (\bibinfo {year} {2017})}\BibitemShut
  {NoStop}%
\bibitem [{\citenamefont {Li}\ \emph {et~al.}(2011)\citenamefont {Li},
  \citenamefont {Walker},\ and\ \citenamefont {Michaelides}}]{li2011quantum}%
  \BibitemOpen
  \bibfield  {author} {\bibinfo {author} {\bibfnamefont {X.-Z.}\ \bibnamefont
  {Li}}, \bibinfo {author} {\bibfnamefont {B.}~\bibnamefont {Walker}}, \ and\
  \bibinfo {author} {\bibfnamefont {A.}~\bibnamefont {Michaelides}},\
  }\href@noop {} {\bibfield  {journal} {\bibinfo  {journal} {Proc. Natl. Acad.
  Sci. U.S.A.}\ }\textbf {\bibinfo {volume} {108}},\ \bibinfo {pages} {6369}
  (\bibinfo {year} {2011})}\BibitemShut {NoStop}%
\bibitem [{\citenamefont {Ceriotti}\ \emph {et~al.}(2013)\citenamefont
  {Ceriotti}, \citenamefont {Cuny}, \citenamefont {Parrinello},\ and\
  \citenamefont {Manolopoulos}}]{ceriotti2013nuclear}%
  \BibitemOpen
  \bibfield  {author} {\bibinfo {author} {\bibfnamefont {M.}~\bibnamefont
  {Ceriotti}}, \bibinfo {author} {\bibfnamefont {J.}~\bibnamefont {Cuny}},
  \bibinfo {author} {\bibfnamefont {M.}~\bibnamefont {Parrinello}}, \ and\
  \bibinfo {author} {\bibfnamefont {D.~E.}\ \bibnamefont {Manolopoulos}},\
  }\href@noop {} {\bibfield  {journal} {\bibinfo  {journal} {Proc. Natl. Acad.
  Sci. U.S.A.}\ }\textbf {\bibinfo {volume} {110}},\ \bibinfo {pages} {15591}
  (\bibinfo {year} {2013})}\BibitemShut {NoStop}%
\bibitem [{\citenamefont {Schreck}\ and\ \citenamefont
  {Wernet}(2016)}]{schreck2016isotope}%
  \BibitemOpen
  \bibfield  {author} {\bibinfo {author} {\bibfnamefont {S.}~\bibnamefont
  {Schreck}}\ and\ \bibinfo {author} {\bibfnamefont {P.}~\bibnamefont
  {Wernet}},\ }\href@noop {} {\bibfield  {journal} {\bibinfo  {journal} {J.
  Chem. Phys.}\ }\textbf {\bibinfo {volume} {145}},\ \bibinfo {pages} {104502}
  (\bibinfo {year} {2016})}\BibitemShut {NoStop}%
\bibitem [{\citenamefont {Harada}\ \emph {et~al.}(2013)\citenamefont {Harada}
  \emph {et~al.}}]{harada2013selective}%
  \BibitemOpen
  \bibfield  {author} {\bibinfo {author} {\bibfnamefont {Y.}~\bibnamefont
  {Harada}} \emph {et~al.},\ }\href@noop {} {\bibfield  {journal} {\bibinfo
  {journal} {Phys. Rev. Lett.}\ }\textbf {\bibinfo {volume} {111}},\ \bibinfo
  {pages} {193001} (\bibinfo {year} {2013})}\BibitemShut {NoStop}%
\bibitem [{\citenamefont {Cisneros}\ \emph {et~al.}(2016)\citenamefont
  {Cisneros} \emph {et~al.}}]{cisneros_ga_2016}%
  \BibitemOpen
  \bibfield  {author} {\bibinfo {author} {\bibfnamefont {G.~A.}\ \bibnamefont
  {Cisneros}} \emph {et~al.},\ }\href@noop {} {\bibfield  {journal} {\bibinfo
  {journal} {Chem. Rev.}\ }\textbf {\bibinfo {volume} {116}},\ \bibinfo {pages}
  {7501} (\bibinfo {year} {2016})}\BibitemShut {NoStop}%
\bibitem [{\citenamefont {Paesani}\ and\ \citenamefont
  {Voth}(2009)}]{paesani_f_2009a}%
  \BibitemOpen
  \bibfield  {author} {\bibinfo {author} {\bibfnamefont {F.}~\bibnamefont
  {Paesani}}\ and\ \bibinfo {author} {\bibfnamefont {G.~A.}\ \bibnamefont
  {Voth}},\ }\href@noop {} {\bibfield  {journal} {\bibinfo  {journal} {J. Phys.
  Chem. B}\ }\textbf {\bibinfo {volume} {113}},\ \bibinfo {pages} {5702}
  (\bibinfo {year} {2009})}\BibitemShut {NoStop}%
\bibitem [{\citenamefont {Babin}\ \emph {et~al.}(2013)\citenamefont {Babin},
  \citenamefont {Leforestier},\ and\ \citenamefont {Paesani}}]{babin_v_2013}%
  \BibitemOpen
  \bibfield  {author} {\bibinfo {author} {\bibfnamefont {V.}~\bibnamefont
  {Babin}}, \bibinfo {author} {\bibfnamefont {C.}~\bibnamefont {Leforestier}},
  \ and\ \bibinfo {author} {\bibfnamefont {F.}~\bibnamefont {Paesani}},\
  }\href@noop {} {\bibfield  {journal} {\bibinfo  {journal} {J. Chem. Theory
  Comput.}\ }\textbf {\bibinfo {volume} {9}},\ \bibinfo {pages} {5395}
  (\bibinfo {year} {2013})}\BibitemShut {NoStop}%
\bibitem [{\citenamefont {Babin}\ \emph {et~al.}(2014)\citenamefont {Babin},
  \citenamefont {Medders},\ and\ \citenamefont {Paesani}}]{babin_v_2014}%
  \BibitemOpen
  \bibfield  {author} {\bibinfo {author} {\bibfnamefont {V.}~\bibnamefont
  {Babin}}, \bibinfo {author} {\bibfnamefont {G.~R.}\ \bibnamefont {Medders}},
  \ and\ \bibinfo {author} {\bibfnamefont {F.}~\bibnamefont {Paesani}},\
  }\href@noop {} {\bibfield  {journal} {\bibinfo  {journal} {J. Chem. Theory
  Comput.}\ }\textbf {\bibinfo {volume} {10}},\ \bibinfo {pages} {1599}
  (\bibinfo {year} {2014})}\BibitemShut {NoStop}%
\bibitem [{\citenamefont {Medders}\ \emph {et~al.}(2014)\citenamefont
  {Medders}, \citenamefont {Babin},\ and\ \citenamefont
  {Paesani}}]{medders_gr_2014}%
  \BibitemOpen
  \bibfield  {author} {\bibinfo {author} {\bibfnamefont {G.~R.}\ \bibnamefont
  {Medders}}, \bibinfo {author} {\bibfnamefont {V.}~\bibnamefont {Babin}}, \
  and\ \bibinfo {author} {\bibfnamefont {F.}~\bibnamefont {Paesani}},\
  }\href@noop {} {\bibfield  {journal} {\bibinfo  {journal} {J. Chem. Theory
  Comput.}\ }\textbf {\bibinfo {volume} {10}},\ \bibinfo {pages} {2906}
  (\bibinfo {year} {2014})}\BibitemShut {NoStop}%
\bibitem [{\citenamefont {Paesani}(2016)}]{paesani2016getting}%
  \BibitemOpen
  \bibfield  {author} {\bibinfo {author} {\bibfnamefont {F.}~\bibnamefont
  {Paesani}},\ }\href@noop {} {\bibfield  {journal} {\bibinfo  {journal} {Acc.
  Chem. Res.}\ }\textbf {\bibinfo {volume} {49}},\ \bibinfo {pages} {1844}
  (\bibinfo {year} {2016})}\BibitemShut {NoStop}%
\bibitem [{\citenamefont {Reddy}\ \emph {et~al.}(2016)\citenamefont {Reddy}
  \emph {et~al.}}]{reddy2016accuracy}%
  \BibitemOpen
  \bibfield  {author} {\bibinfo {author} {\bibfnamefont {S.~K.}\ \bibnamefont
  {Reddy}} \emph {et~al.},\ }\href@noop {} {\bibfield  {journal} {\bibinfo
  {journal} {J. Chem. Phys.}\ }\textbf {\bibinfo {volume} {145}},\ \bibinfo
  {pages} {194504} (\bibinfo {year} {2016})}\BibitemShut {NoStop}%
\bibitem [{\citenamefont {Brown}\ \emph {et~al.}(2017)\citenamefont {Brown},
  \citenamefont {G\"{o}tz}, \citenamefont {Cheng}, \citenamefont {Steele},
  \citenamefont {Mandelshtam},\ and\ \citenamefont {Paesani}}]{brown_se_2017}%
  \BibitemOpen
  \bibfield  {author} {\bibinfo {author} {\bibfnamefont {S.~E.}\ \bibnamefont
  {Brown}}, \bibinfo {author} {\bibfnamefont {A.~W.}\ \bibnamefont {G\"{o}tz}},
  \bibinfo {author} {\bibfnamefont {X.}~\bibnamefont {Cheng}}, \bibinfo
  {author} {\bibfnamefont {R.~P.}\ \bibnamefont {Steele}}, \bibinfo {author}
  {\bibfnamefont {V.~A.}\ \bibnamefont {Mandelshtam}}, \ and\ \bibinfo {author}
  {\bibfnamefont {F.}~\bibnamefont {Paesani}},\ }\href@noop {} {\bibfield
  {journal} {\bibinfo  {journal} {J. Am. Chem. Soc.}\ }\textbf {\bibinfo
  {volume} {139}},\ \bibinfo {pages} {7082} (\bibinfo {year}
  {2017})}\BibitemShut {NoStop}%
\bibitem [{\citenamefont {Reddy}\ \emph
  {et~al.}(2017{\natexlab{a}})\citenamefont {Reddy}, \citenamefont {Moberg},
  \citenamefont {Straight},\ and\ \citenamefont {Paesani}}]{reddy_sk_2017}%
  \BibitemOpen
  \bibfield  {author} {\bibinfo {author} {\bibfnamefont {S.~K.}\ \bibnamefont
  {Reddy}}, \bibinfo {author} {\bibfnamefont {D.~R.}\ \bibnamefont {Moberg}},
  \bibinfo {author} {\bibfnamefont {S.~C.}\ \bibnamefont {Straight}}, \ and\
  \bibinfo {author} {\bibfnamefont {F.}~\bibnamefont {Paesani}},\ }\href@noop
  {} {\bibfield  {journal} {\bibinfo  {journal} {J. Chem. Phys.}\ }\textbf
  {\bibinfo {volume} {147}} (\bibinfo {year} {2017}{\natexlab{a}})}\BibitemShut
  {NoStop}%
\bibitem [{\citenamefont {Pham}\ \emph {et~al.}(2017)\citenamefont {Pham},
  \citenamefont {Reddy}, \citenamefont {Chen}, \citenamefont {Knight},\ and\
  \citenamefont {Paesani}}]{pham_ch_2017}%
  \BibitemOpen
  \bibfield  {author} {\bibinfo {author} {\bibfnamefont {C.~H.}\ \bibnamefont
  {Pham}}, \bibinfo {author} {\bibfnamefont {S.~K.}\ \bibnamefont {Reddy}},
  \bibinfo {author} {\bibfnamefont {K.}~\bibnamefont {Chen}}, \bibinfo {author}
  {\bibfnamefont {C.}~\bibnamefont {Knight}}, \ and\ \bibinfo {author}
  {\bibfnamefont {F.}~\bibnamefont {Paesani}},\ }\href {\doibase
  10.1021/acs.jctc.6b01248} {\bibfield  {journal} {\bibinfo  {journal} {J.
  Chem. Theory Comput.}\ }\textbf {\bibinfo {volume} {13}},\ \bibinfo {pages}
  {1778} (\bibinfo {year} {2017})}\BibitemShut {NoStop}%
\bibitem [{\citenamefont {Moberg}\ \emph {et~al.}(2017)\citenamefont {Moberg},
  \citenamefont {Straight}, \citenamefont {Knight},\ and\ \citenamefont
  {Paesani}}]{moberg_dr_2017}%
  \BibitemOpen
  \bibfield  {author} {\bibinfo {author} {\bibfnamefont {D.~R.}\ \bibnamefont
  {Moberg}}, \bibinfo {author} {\bibfnamefont {S.~C.}\ \bibnamefont
  {Straight}}, \bibinfo {author} {\bibfnamefont {C.}~\bibnamefont {Knight}}, \
  and\ \bibinfo {author} {\bibfnamefont {F.}~\bibnamefont {Paesani}},\
  }\href@noop {} {\bibfield  {journal} {\bibinfo  {journal} {J. Phys. Chem.
  Lett.}\ }\textbf {\bibinfo {volume} {8}},\ \bibinfo {pages} {2579} (\bibinfo
  {year} {2017})}\BibitemShut {NoStop}%
\bibitem [{\citenamefont {Kang}\ and\ \citenamefont
  {Hybertsen}(2010)}]{kang2010enhanced}%
  \BibitemOpen
  \bibfield  {author} {\bibinfo {author} {\bibfnamefont {W.}~\bibnamefont
  {Kang}}\ and\ \bibinfo {author} {\bibfnamefont {M.~S.}\ \bibnamefont
  {Hybertsen}},\ }\href@noop {} {\bibfield  {journal} {\bibinfo  {journal}
  {Phys. Rev. B}\ }\textbf {\bibinfo {volume} {82}},\ \bibinfo {pages} {195108}
  (\bibinfo {year} {2010})}\BibitemShut {NoStop}%
\bibitem [{\citenamefont {Wu}\ \emph {et~al.}(2009)\citenamefont {Wu},
  \citenamefont {Selloni},\ and\ \citenamefont {Car}}]{09B-Wu}%
  \BibitemOpen
  \bibfield  {author} {\bibinfo {author} {\bibfnamefont {X.}~\bibnamefont
  {Wu}}, \bibinfo {author} {\bibfnamefont {A.}~\bibnamefont {Selloni}}, \ and\
  \bibinfo {author} {\bibfnamefont {R.}~\bibnamefont {Car}},\ }\href {\doibase
  10.1103/PhysRevB.79.085102} {\bibfield  {journal} {\bibinfo  {journal} {Phys.
  Rev. B}\ }\textbf {\bibinfo {volume} {79}},\ \bibinfo {pages} {085102}
  (\bibinfo {year} {2009})}\BibitemShut {NoStop}%
\bibitem [{\citenamefont {Giannozzi}\ and\ \citenamefont
  {et~al.}(2017)}]{giannozzi2017advanced}%
  \BibitemOpen
  \bibfield  {author} {\bibinfo {author} {\bibfnamefont {P.}~\bibnamefont
  {Giannozzi}}\ and\ \bibinfo {author} {\bibnamefont {et~al.}},\ }\href@noop {}
  {\bibfield  {journal} {\bibinfo  {journal} {J. Phys. Condens. Matter}\
  }\textbf {\bibinfo {volume} {29}},\ \bibinfo {pages} {465901} (\bibinfo
  {year} {2017})}\BibitemShut {NoStop}%
\bibitem [{\citenamefont {Martyna}\ \emph {et~al.}(1992)\citenamefont
  {Martyna}, \citenamefont {Klein},\ and\ \citenamefont
  {Tuckerman}}]{Nose1992}%
  \BibitemOpen
  \bibfield  {author} {\bibinfo {author} {\bibfnamefont {G.~J.}\ \bibnamefont
  {Martyna}}, \bibinfo {author} {\bibfnamefont {M.~L.}\ \bibnamefont {Klein}},
  \ and\ \bibinfo {author} {\bibfnamefont {M.}~\bibnamefont {Tuckerman}},\
  }\href@noop {} {\bibfield  {journal} {\bibinfo  {journal} {J. Chem. Phys.}\
  }\textbf {\bibinfo {volume} {97}},\ \bibinfo {pages} {2635} (\bibinfo {year}
  {1992})}\BibitemShut {NoStop}%
\bibitem [{\citenamefont {Berne}\ and\ \citenamefont
  {Thirumalai}(1986)}]{berne1986}%
  \BibitemOpen
  \bibfield  {author} {\bibinfo {author} {\bibfnamefont {B.~J.}\ \bibnamefont
  {Berne}}\ and\ \bibinfo {author} {\bibfnamefont {D.}~\bibnamefont
  {Thirumalai}},\ }\href@noop {} {\bibfield  {journal} {\bibinfo  {journal}
  {Ann. Rev. Phys. Chem.}\ }\textbf {\bibinfo {volume} {37}},\ \bibinfo {pages}
  {401} (\bibinfo {year} {1986})}\BibitemShut {NoStop}%
\bibitem [{\citenamefont {Reddy}\ \emph
  {et~al.}(2017{\natexlab{b}})\citenamefont {Reddy}, \citenamefont {Moberg},
  \citenamefont {Straight},\ and\ \citenamefont {Paesani}}]{reddy2017}%
  \BibitemOpen
  \bibfield  {author} {\bibinfo {author} {\bibfnamefont {S.~K.}\ \bibnamefont
  {Reddy}}, \bibinfo {author} {\bibfnamefont {D.~R.}\ \bibnamefont {Moberg}},
  \bibinfo {author} {\bibfnamefont {S.~C.}\ \bibnamefont {Straight}}, \ and\
  \bibinfo {author} {\bibfnamefont {F.}~\bibnamefont {Paesani}},\ }\href
  {\doibase 10.1063/1.5006480} {\bibfield  {journal} {\bibinfo  {journal} {J.
  Chem. Phys.}\ }\textbf {\bibinfo {volume} {147}},\ \bibinfo {pages} {244504}
  (\bibinfo {year} {2017}{\natexlab{b}})}\BibitemShut {NoStop}%
\bibitem [{\citenamefont {Ceriotti}\ and\ \citenamefont
  {Manolopoulos}(2012)}]{ceriotti2012efficient}%
  \BibitemOpen
  \bibfield  {author} {\bibinfo {author} {\bibfnamefont {M.}~\bibnamefont
  {Ceriotti}}\ and\ \bibinfo {author} {\bibfnamefont {D.~E.}\ \bibnamefont
  {Manolopoulos}},\ }\href@noop {} {\bibfield  {journal} {\bibinfo  {journal}
  {Phys. Rev. Lett.}\ }\textbf {\bibinfo {volume} {109}},\ \bibinfo {pages}
  {100604} (\bibinfo {year} {2012})}\BibitemShut {NoStop}%
\bibitem [{\citenamefont {Onida}\ \emph {et~al.}(2002)\citenamefont {Onida},
  \citenamefont {Reining},\ and\ \citenamefont {Rubio}}]{onida2002electronic}%
  \BibitemOpen
  \bibfield  {author} {\bibinfo {author} {\bibfnamefont {G.}~\bibnamefont
  {Onida}}, \bibinfo {author} {\bibfnamefont {L.}~\bibnamefont {Reining}}, \
  and\ \bibinfo {author} {\bibfnamefont {A.}~\bibnamefont {Rubio}},\
  }\href@noop {} {\bibfield  {journal} {\bibinfo  {journal} {Rev. Mod. Phys.}\
  }\textbf {\bibinfo {volume} {74}},\ \bibinfo {pages} {601} (\bibinfo {year}
  {2002})}\BibitemShut {NoStop}%
\bibitem [{\citenamefont {K{\"u}hne}\ and\ \citenamefont
  {Khaliullin}(2013)}]{kuhne2013electronic}%
  \BibitemOpen
  \bibfield  {author} {\bibinfo {author} {\bibfnamefont {T.~D.}\ \bibnamefont
  {K{\"u}hne}}\ and\ \bibinfo {author} {\bibfnamefont {R.~Z.}\ \bibnamefont
  {Khaliullin}},\ }\href@noop {} {\bibfield  {journal} {\bibinfo  {journal}
  {Nat. Commun.}\ }\textbf {\bibinfo {volume} {4}},\ \bibinfo {pages} {1450}
  (\bibinfo {year} {2013})}\BibitemShut {NoStop}%
\bibitem [{\citenamefont {Hassanali}\ \emph {et~al.}(2013)\citenamefont
  {Hassanali}, \citenamefont {Giberti}, \citenamefont {Cuny}, \citenamefont
  {K{\"u}hne},\ and\ \citenamefont {Parrinello}}]{hassanali2013proton}%
  \BibitemOpen
  \bibfield  {author} {\bibinfo {author} {\bibfnamefont {A.}~\bibnamefont
  {Hassanali}}, \bibinfo {author} {\bibfnamefont {F.}~\bibnamefont {Giberti}},
  \bibinfo {author} {\bibfnamefont {J.}~\bibnamefont {Cuny}}, \bibinfo {author}
  {\bibfnamefont {T.~D.}\ \bibnamefont {K{\"u}hne}}, \ and\ \bibinfo {author}
  {\bibfnamefont {M.}~\bibnamefont {Parrinello}},\ }\href@noop {} {\bibfield
  {journal} {\bibinfo  {journal} {Proc. Natl. Acad. Sci. U.S.A.}\ }\textbf
  {\bibinfo {volume} {110}},\ \bibinfo {pages} {13723} (\bibinfo {year}
  {2013})}\BibitemShut {NoStop}%
\bibitem [{\citenamefont {Hart}\ \emph {et~al.}(2005)\citenamefont {Hart},
  \citenamefont {Benmore}, \citenamefont {Neuefeind}, \citenamefont {Kohara},
  \citenamefont {Tomberli},\ and\ \citenamefont
  {Egelstaff}}]{hart2005temperature}%
  \BibitemOpen
  \bibfield  {author} {\bibinfo {author} {\bibfnamefont {R.}~\bibnamefont
  {Hart}}, \bibinfo {author} {\bibfnamefont {C.}~\bibnamefont {Benmore}},
  \bibinfo {author} {\bibfnamefont {J.}~\bibnamefont {Neuefeind}}, \bibinfo
  {author} {\bibfnamefont {S.}~\bibnamefont {Kohara}}, \bibinfo {author}
  {\bibfnamefont {B.}~\bibnamefont {Tomberli}}, \ and\ \bibinfo {author}
  {\bibfnamefont {P.}~\bibnamefont {Egelstaff}},\ }\href@noop {} {\bibfield
  {journal} {\bibinfo  {journal} {Phys. Rev. Lett.}\ }\textbf {\bibinfo
  {volume} {94}},\ \bibinfo {pages} {047801} (\bibinfo {year}
  {2005})}\BibitemShut {NoStop}%
\bibitem [{\citenamefont {Hart}\ \emph {et~al.}(2006)\citenamefont {Hart} \emph
  {et~al.}}]{hart2006isotope}%
  \BibitemOpen
  \bibfield  {author} {\bibinfo {author} {\bibfnamefont {R.}~\bibnamefont
  {Hart}} \emph {et~al.},\ }\href@noop {} {\bibfield  {journal} {\bibinfo
  {journal} {J. Chem. Phys.}\ }\textbf {\bibinfo {volume} {124}},\ \bibinfo
  {pages} {134505} (\bibinfo {year} {2006})}\BibitemShut {NoStop}%
\bibitem [{\citenamefont {Gallo}\ \emph {et~al.}(2016)\citenamefont {Gallo}
  \emph {et~al.}}]{gallo2016water}%
  \BibitemOpen
  \bibfield  {author} {\bibinfo {author} {\bibfnamefont {P.}~\bibnamefont
  {Gallo}} \emph {et~al.},\ }\href@noop {} {\bibfield  {journal} {\bibinfo
  {journal} {Chem. Rev.}\ }\textbf {\bibinfo {volume} {116}},\ \bibinfo {pages}
  {7463} (\bibinfo {year} {2016})}\BibitemShut {NoStop}%
\end{thebibliography}

%

\end{document}